\documentclass{iopart}
\usepackage{graphicx}
\begin{document}
\title{Relation between color deconfinement and chiral restoration
 at finite temperature and density}
\author{Kenji Fukushima}
\address{Center for Theoretical Physics,
         Massachusetts Institute of Technology,\\
         77 Massachusetts Avenue, Cambridge, MA 02139, U.S.A.}
\address{Department of Physics,
         University of Tokyo,\\
         7-3-1 Hongo, Bunkyo-ku, Tokyo 113-0033, Japan}
\begin{abstract}
We discuss the relation between color deconfinement characterized by
the Polyakov loop and chiral restoration characterized by the chiral
condensate. We clarify how and why these two phenomena should take
place at the same pseudo-critical temperature.
\end{abstract}

Quantum Chromodynamics (QCD) is believed to undergo phase transitions
from hadronic matter to a quark-gluon plasma (QGP) at high
temperature. It is well known that the QGP phase transition can be
exactly defined only in the extreme cases, namely, the heavy quark
limit ($m_\mathrm{q}=0$) corresponding to deconfinement and the chiral
limit ($m_\mathrm{q}=0$) corresponding to chiral restoration. The
question is then; what would the QCD phase transition be like for an
intermediate value of $m_\mathrm{q}$? Our conclusion here is that
there is \textit{only one phenomenon} for any $m_\mathrm{q}$, that is
an admixture of deconfinement and chiral restoration.

The color confinement-deconfinement phase transition at finite
temperature is diagnosed by means of an order parameter called the
Polyakov loop. It is essentially the exponential of the free energy
gain with a quark placed in a thermal medium, i.e.,
$l\sim\exp[-f_0/T]$. If the vacuum is color confining, $f_0$ diverges
leading to $l=0$, while $l$ takes a finite value in the deconfined
phase. This criterion, however, does not work when dynamical quarks in
the fundamental representation in color space are thermally excited.
Then $f_0$ no longer diverges because the color-flux tube can break
apart with mesonic excitation energies. The chiral phase transition
is, on the other hand, manifested in case of massless quarks. The
order parameter is given by a condensate in the sigma-meson channel,
namely, the chiral condensate $\langle\bar{q}q\rangle$. A finite quark
mass would blur a sharp phase transition.

\begin{figure}
\begin{center}
\includegraphics[width=4cm]{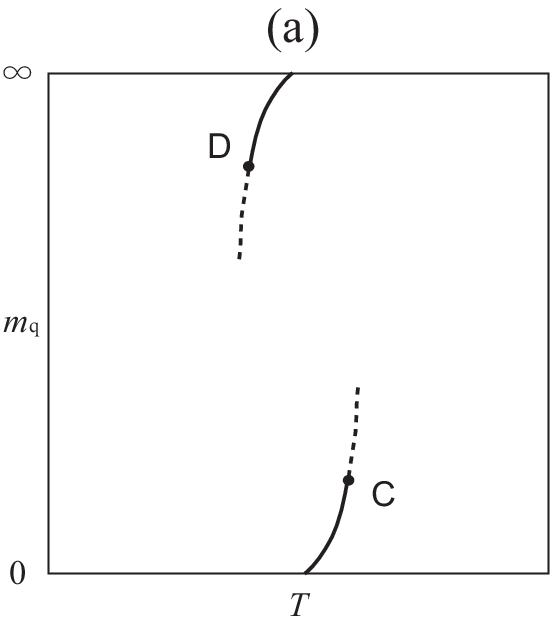}\hspace{1cm}
\includegraphics[width=4cm]{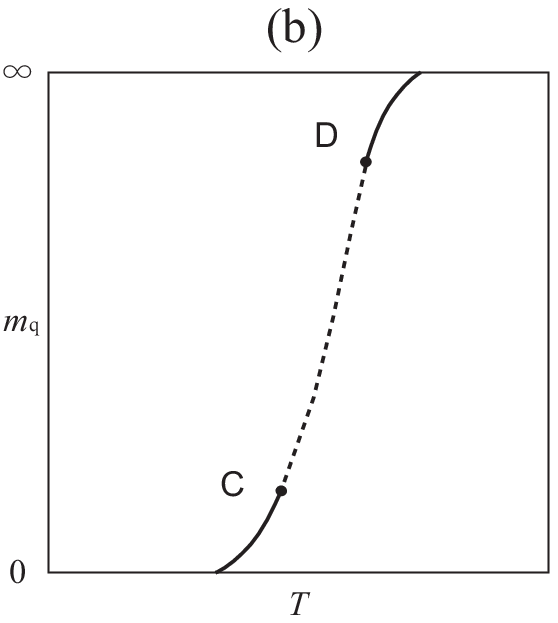}
\caption{Schematic QCD phase diagrams in the ($T$,\,$m_\mathrm{q}$)
plane. Solid curves represent the first order phase boundary and the
dotted curves represent the continuous crossover.}
\label{fig:phase}
\end{center}
\end{figure}

It is widely accepted that the order of phase transitions is as
follows. The deconfinement transition with $m_\mathrm{q}=\infty$ is of
second order for $N_\mathrm{c}=2$ and of first order for
$N_\mathrm{c}\ge 3$, and the chiral phase transition with
$m_\mathrm{q}=0$ is of second order for $N_\mathrm{f}=2$ and of first
order for $N_\mathrm{f}\ge 3$ where $N_\mathrm{c}$ and $N_\mathrm{f}$
represent the color and flavor numbers respectively. Thus in the QCD
case with $N_\mathrm{c}=3$ and $N_\mathrm{f}\sim3$ the phase
transitions are both of first order starting from the top (heavy quark
limit) and the bottom (chiral limit) edges in
Fig.~\ref{fig:phase}~(a).
They end with second order critical end-points (\textsf{C} and
\textsf{D}) and the rests are crossovers where the peak in
susceptibilities yields the pseudo-critical temperature. One may well
naively draw a phase diagram in the ($T$,\,$m_q$) plane like
Fig.~\ref{fig:phase}~(a). If this were true, the deconfinement and
chiral phase transitions would lie only in different regions with
respect to the quark mass and have little to do with each other.

The lattice QCD data \cite{fuk86,kar94,aok98}, however, strongly
suggest that two end-points, \textsf{C} and \textsf{D}, are connected
by a single crossover line as shown in Fig.~\ref{fig:phase}~(b). The
evidence comes from the following observations; 1) The Polyakov loop
and chiral susceptibilities have only \textit{one} peak as functions
of the temperature and their locations (pseudo-critical temperatures)
exactly coincide with each other. This coincidence is seen for
\textit{any} value of $m_\mathrm{q}$. 2) At the pseudo-critical
temperature inferred from the susceptibility peak, the energy density
rapidly increases, which signifies the liberation of color degrees of
freedom.  This means that, even if $m_\mathrm{q}$ is not so large, the
QCD phase transition should have a substantial remnant of
deconfinement in contradiction to a naive picture depicted in
Fig.~\ref{fig:phase}~(a).

It is often said that QCD with realistic quark masses has only one
(approximate) phase transition, namely, the chiral phase transition,
and the coincidence of the pseudo-critical temperatures is a
consequence of the mixing between the Polyakov loop and the chiral
condensate \cite{goc85,sat98,dig01,fuk03_1,san03}. This statement is
half-right and half-wrong. Certainly there is only one phase
transition, but it should convey the information on deconfinement as
well as chiral restoration. The energy density measured on the
lattice, otherwise, could not have enough explanations and even the
experimental interpretation on the QGP would become suspicious. In
fact, since the deconfinement and chiral phase transitions are only
different manifestations of the \textit{same phenomena}, as we
emphasize here, the notion of the QGP can have a definite meaning.

\begin{figure}
\begin{center}
\includegraphics[width=4cm]{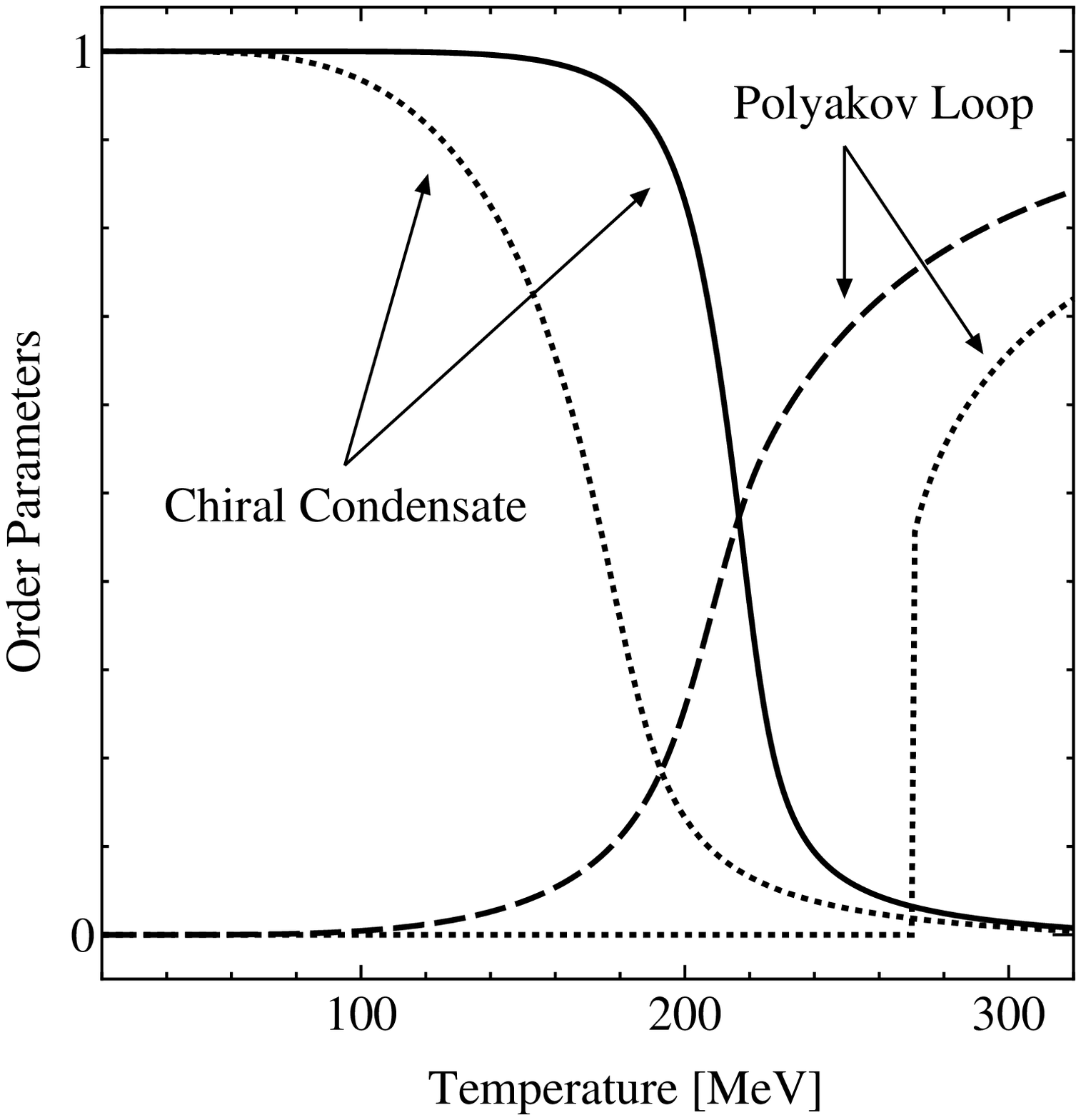}
\includegraphics[width=4cm]{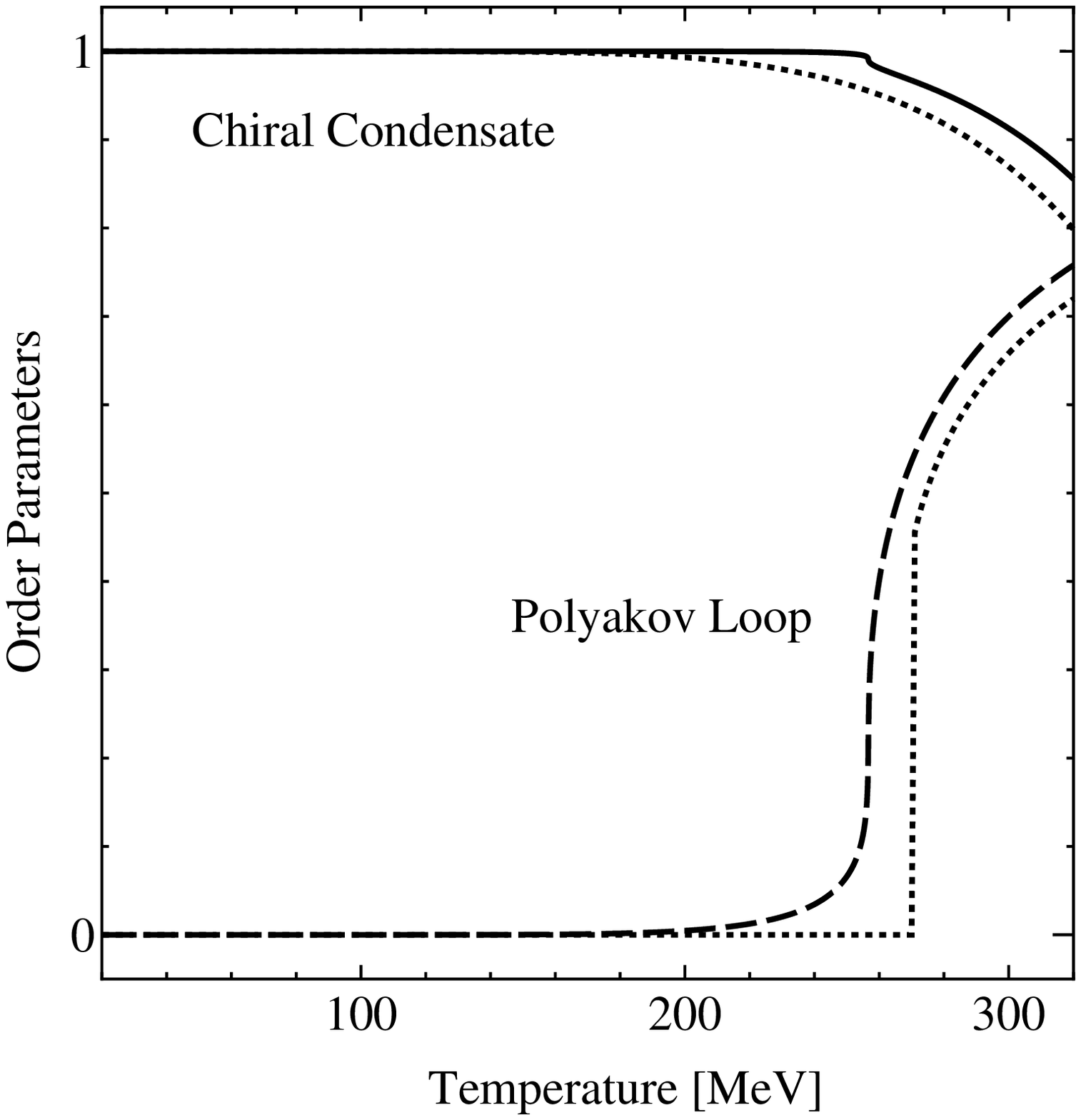}
\includegraphics[width=4cm]{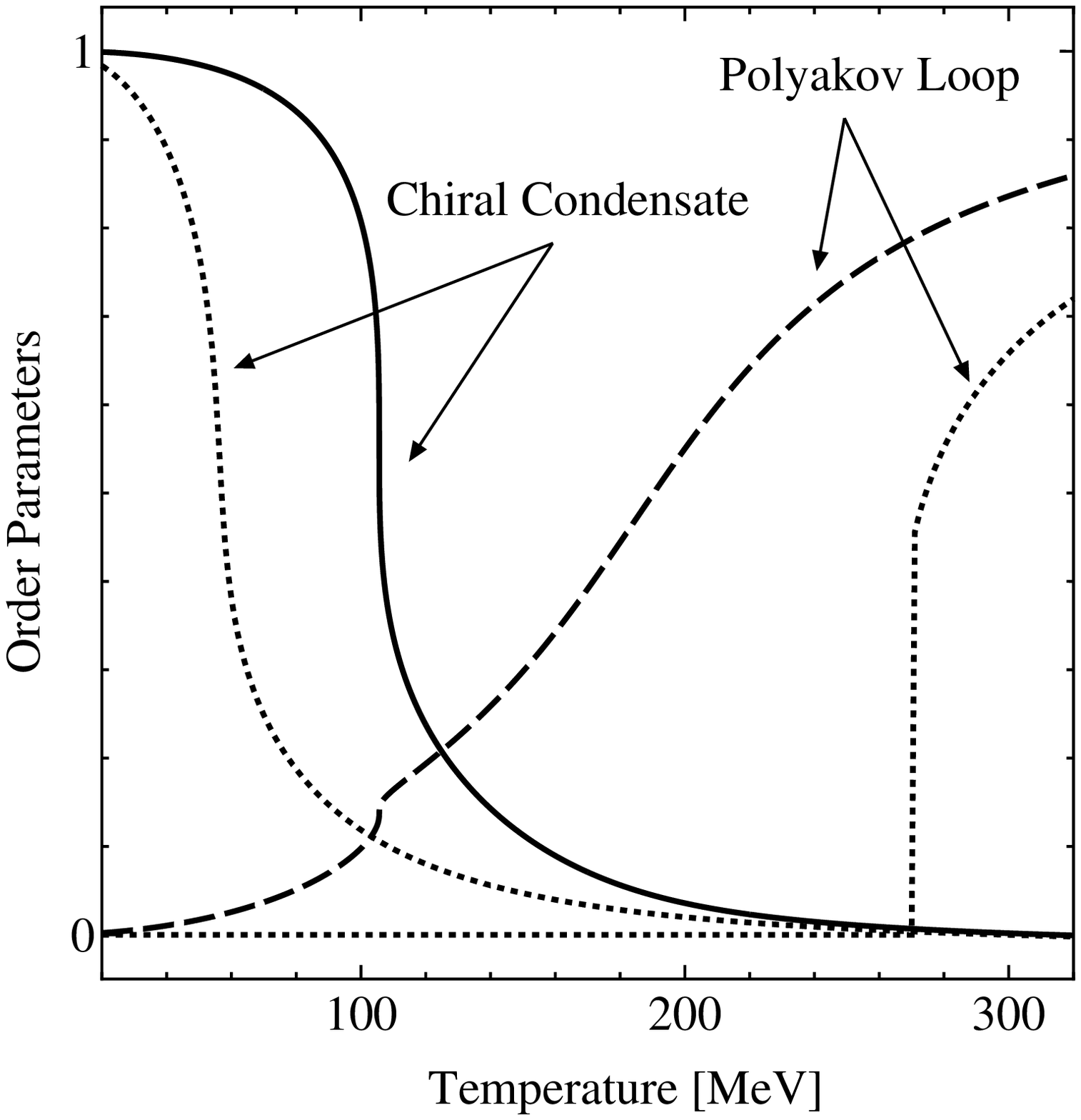}
\caption{The behavior of order parameters with the interaction term
(\ref{eq:int}). The dashed curves are the results without any
interaction between the order parameters. The left figure shows the
Polyakov loop and the chiral condensate for
$m_\mathrm{q}=5.5\,\mbox{MeV}$ and $\mu_\mathrm{q}=0$. The middle
figure represents the results at \textsf{D} with
$m_\mathrm{q}=788\,\mbox{MeV}$ and $\mu_\mathrm{q}=0$. The right
figure shows the results at another end-point at finite density with
$m_\mathrm{q}=5.5\,\mbox{MeV}$ and $\mu_\mathrm{q}=312\,\mbox{MeV}$.}
\label{fig:model}
\end{center}
\end{figure}

Now we shall discuss how the simultaneous crossovers occur
\cite{fuk03_2}. In principle the effective potential in terms of the
Polyakov loop, $l$, and the chiral condensate,
$\langle\bar{q}q\rangle$, would tell us everything about the phase
structure. The principle to write down the effective potential is the
center symmetry and the chiral symmetry, both of which are explicitly
broken by a non-diverging and non-vanishing quark mass. From the
experience in the hopping parameter expansion on the lattice we can
consider that the center symmetry breaking term generally appears as a
result of the coupling between the Polyakov loop and the thermal
(temporal) quark propagation, i.e., $l\cdot\exp[-E_\mathrm{q}/T]$
where $E_\mathrm{q}$ is the quasi-quark energy depending on
$\langle\bar{q}q\rangle$ through the constituent quark mass. The
simplest choice of the interaction term is
\begin{equation}
 \sim \int\frac{\mathrm{d}^3p}{(2\pi)^3}\mathrm{Tr_c}\Bigl\{
  \ln\bigl[1+L\mathrm{e}^{-(E_\mathrm{q}-\mu_\mathrm{q})/T}\bigr]
  +\ln\bigl[1+L^\dagger\mathrm{e}^{-(E_\mathrm{q}+\mu_\mathrm{q})/T}
  \bigr]\Bigl\},
\label{eq:int}
\end{equation}
where $L$ is the untraced (SU(3) matrix) Polyakov loop and
$\mu_\mathrm{q}$ is the quark chemical potential. In
Fig.~\ref{fig:model} we present typical numerical outputs from a
chiral effective model with the interaction term (\ref{eq:int}).

The left figure shows the results for $m_\mathrm{q}=5.5\,\mbox{MeV}$ at
zero quark density. It is obvious that two crossovers occur almost
simultaneously as a result of the interaction. The reason is easy to
understand. As long as the expectation value of the Polyakov loop is
small, the Polyakov loop matrix takes any direction in color space.
Since the  quark excitation is accompanied by the Polyakov loop in
(\ref{eq:int}), averaging over the color directions leads to colorless
excitations made of composite quarks. Then thermal excitations are
suppressed by twice (for meson-like excitations) or thrice (for
baryon-like excitations) larger excitation energies. Once the Polyakov
loop comes to take a finite expectation value, dynamical quarks
participate in thermal excitations, which makes the chiral condensate
decrease simultaneously. The right figure is the model prediction for
the behavior at finite density. Since the center symmetry is
explicitly broken by the density effect, the Polyakov loop is
significantly affected to have a smoother slope.

This model works well for small $m_\mathrm{q}$ in particular. The
interaction term (\ref{eq:int}) is not everything, however. The phase
diagram shown in Fig.~\ref{fig:phase}~(b) needs a stronger constraint.
A correct interpretation on the phase diagram can be provided by the
sigma-glueball mixing in a transparent manner \cite{hat03}. The
Polyakov loop lacks a physical interpretation in Minkowskian
space-time. Instead, we can make use of the electric glueball,
$G_\mathrm{E}$, to characterize the deconfinement phase transition.
Roughly speaking, the electric glueball containing the electric
(temporal) component of the gluon fields directly couples to the
Polyakov loop at finite temperature, and thus we can regard the
electric glueball as almost the same as the Polyakov loop near the
critical point. As is well known in the excitation spectrum at zero
temperature, there is generally a finite mixing between $\sigma$ and
$G_\mathrm{E}$. As a result of the mixing we have two eigen-fields,
\begin{equation}
 \phi = \sigma\cos\theta + G_\mathrm{E}\sin\theta,\quad
 \phi' = -\sigma\sin\theta + G_\mathrm{E}\cos\theta,
\end{equation}
where $\sigma=\bar{q}q$ and $G_\mathrm{E}=|l-\langle l\rangle|^2$.

Fig.~\ref{fig:phase}~(b) implies that only the $\phi$ field should be
light enough and the heavy $\phi'$ field decouples in the vicinity of
the critical point. The order parameter field is almost $\sigma$ and
$G_\mathrm{E}$ at \textsf{C} and \textsf{D} respectively, and the
mixing angle continuously changes from $\theta\sim0$ to
$\theta\sim\pi/2$ between \textsf{C} and \textsf{D}. This nature of
the QCD phase transition can be confirmed by the lattice QCD
simulation. The measurement of the screening lengths in the $\sigma$
and $G_\mathrm{E}$ coupled channel, for instance, would give the
information on $\theta$ as a function of $m_\mathrm{q}$.

The reason why only the $\phi$ field can stay light is to be
understood by the \textit{level repulsion} between the $\phi$ and
$\phi'$ states. If the level repulsion is strong enough, one field
($\phi$ field) is pushed down to be light and the other ($\phi'$
field) is pushed up to be heavy. In fact, however, the level repulsion
always occurs however small the mixing interaction is. Therefore the
level repulsion itself cannot be a sufficient condition to account for
the QCD phase diagram. Nevertheless, such an intuitive understanding
based on the level repulsion would be useful to take an overview at
the gross feature. How strong the level repulsion is should be
revealed in the future lattice QCD simulation.

From the point of view of the level repulsion the model study with the
interaction term (\ref{eq:int}) cannot yield satisfactory results for
large $m_\mathrm{q}$. Even though the simultaneous crossovers occur,
the mixing is not strong enough to lead to a level repulsion. Then the
level crossing makes the susceptibility with undesirable two peaks for
large $m_\mathrm{q}$. What realizes a strong level repulsion in
reality and the perfect locking between deconfinement and chiral
restoration is still an open question.

In summary, we discussed the QCD phase transition. We pointed out the
correct interpretation on the lattice QCD data. The most important
is that the lattice data signify much more than a simple mixing
argument between the sigma and the Polyakov loop. We showed the
results in model analyses to demonstrate how the simultaneous
crossovers could occur. The model study turned out useful to specify
the important property in the dynamics, but not sufficient to give an
explanation on the whole phase diagram. We also expressed the
interpretation on the QCD phase diagram in a transparent manner by
using the sigma-glueball mixing. The future lattice calculation should
elucidate quantitative details in our scenario.

Y.~Hatta is acknowledged for stimulating collaboration. This work is
supported in part by Research Fellowships of the Japan Society for the
Promotion of Science for Young Scientists and by funds provided by the
U.S. Department of Energy (D.O.E.) under cooperative research
agreement \#DF-FC02-94ER40818.\\


\begin{thebibliography}{99}

\bibitem{fuk86}
 M. Fukugita and A. Ukawa, Phys. Rev. Lett. 57 (1986) 503.
\bibitem{kar94}
 F. Karsch and E. Laermann, Phys. Rev. D50 (1994) 6954.
\bibitem{aok98}
 S. Aoki, M. Fukugita, S. Hashimoto, N. Ishizuka, Y. Iwasaki,
 K. Kanaya, K. Kuramashi, H. Mino, M. Okawa, A. Ukawa and
 T. Yoshi\'{e}, Phys. Rev. D57 (1998) 3910.
\bibitem{goc85}
 A. Gocksch and M. Ogilvie, Phys. Rev. D31 (1985) 877.
\bibitem{sat98}
 H. Satz, Nucl. Phys. A642 (1998) 130.
\bibitem{dig01}
 S. Digal, E. Laermann and H. Satz, Eur. Phys. J. C18 (2001) 583.
\bibitem{fuk03_1}
 K. Fukushima, Phys. Lett. B553 (2003) 38.
\bibitem{san03}
 A. Mocsy, F. Sannino and K. Tuominen,
  e-print: \texttt{hep-ph/0308135}.
\bibitem{fuk03_2}
 K. Fukushima, Phys. Rev. D68 (2003) 045004;
  e-print: \texttt{hep-ph/0310121}.
\bibitem{hat03}
 Y. Hatta and K. Fukushima, e-print:
  \texttt{hep-ph/0307068} to appear in Phys. Rev. D;
  e-print: \texttt{hep-ph/0311267}.
\end{thebibliography}
\end{document}